\pdfoutput=1
\documentclass[preprint]{aastex}
\usepackage{natbib}

\usepackage{amsmath,amstext}

\shorttitle{Space Weather Environment of Proxima b}
\shortauthors{C. Garraffo et al.}

\begin{document}

\title{THE SPACE WEATHER OF PROXIMA CENTAURI b}
\author{Garraffo, C.\altaffilmark{1}, Drake, J.J.\altaffilmark{1}, Cohen, O.\altaffilmark{2,1}}
\email{cgarraffo@cfa.harvard.edu}
\altaffiltext{1}{Harvard-Smithsonian Center for Astrophysics, 60 Garden Street, Cambridge MA 02138}
\altaffiltext{2}{Lowell Center for Space Science and Technology, University of Massachusetts, Lowell, MA 01854, USA}

\begin{abstract}

A planet orbiting in the "habitable zone" of our closest neighboring star, Proxima Centauri, has recently been  discovered, and the next natural question is whether or not Proxima~b is``habitable".
Stellar winds are likely a source of atmospheric erosion that could be particularly severe in the case of M dwarf habitable zone planets that reside close to their parent star. Here we study the stellar wind conditions that Proxima~b experiences over its orbit.
We construct 3-D MHD models of the wind and magnetic field 
around Proxima Centauri using a surface magnetic field map for a star of the same spectral type and scaled to match the observed $\sim 600$~G surface magnetic field strength of Proxima.
We examine the wind conditions and dynamic pressure over different plausible orbits that sample the constrained parameters of  the orbit of Proxima~b. For all the parameter space explored, the planet is subject to stellar wind pressures of more than 2000 times those experienced by Earth from the solar wind.  During an orbit, Proxima~b is also subject to pressure changes of 1 to 3 orders of magnitude on timescales of a day. Its magnetopause standoff distance consequently undergoes sudden and periodic changes by a factor of 2 to 5. Proxima~b will traverse the interplanetary current sheet twice each orbit, and likely crosses into regions of subsonic wind quite frequently. These effects should be taken into account in any physically realistic assessment or prediction of its atmospheric reservoir, characteristics and loss.\\

\end{abstract}


\section{Introduction}
\label{s:intro}

The recent discovery of the planet Proxima~b orbiting in the nominal ``habitable zone" of our nearest stellar neighbor \citep{Anglada-Escude.etal:16} presents a unique opportunity to further study and infer the properties and the evolutionary path of exoplanets.   It has been pointed out that it is sufficiently close to Earth as to be directly observable by the next generation of space telescopes such as WFIRST and JWST, in addition to planned 30-meter class ground-based telescopes \citep{Barnes.etal:16,Kreidberg.Loeb:16}.

Proxima b is estimated to be of at least 1.3 Earth masses ($M\sin i=1.3 M_\Earth$) and have an orbital period of 11.2 days with a semi-major axis of only 0.049~AU---twenty times closer to Proxima than the Earth is to the Sun  \citep{Anglada-Escude.etal:16}.  Apart from its approximate mass and orbital parameters, rather little is currently known about Proxima~b itself, although a handful of studies have already examined its likely irradiation history and possible climate and evolution in relation to potential habitability \citep{Ribas.etal:16,Turbet.etal:16,Barnes.etal:16,Meadows.etal:16}.

The current view of the ``habitable" zone of a star is limited to a rather narrow definition of the range of orbital distances over which a planet might have liquid surface water (e.g., \citealt{Kasting.etal:93}). Of special additional importance for planets around M dwarfs such as Proxima (M5.5V) is the potential for such a habitable zone planet to retain any surface atmosphere at all over a sufficiently long period of time that renders habitability of practical interest.  M dwarfs are potentially awkward places for atmospheres to survive because their UV, EUV and X-ray 
(hereafter UEX\footnote{We purposely avoid the acronym XUV that has sometimes been adopted as a shorthand for the UV to X-ray bands due to potential confusion with its common historical use to describe the extreme ultraviolet band covering approximately 100 to 912~\AA.}) 
radiation that can drive atmospheric photoevaporation \citep[e.g.][]{Lammer.etal:03,Penz.Micela:08,Owen.Jackson:12} stays proportionally larger in relation to their bolometric luminosity for much longer than higher mass Sun-like stars \citep[e.g.][]{Wright.etal:11,Jackson.etal:12}.

While photoevaporation of planetary atmospheres due to parent stellar radiation has been relatively well-studied in limited regimes, the stellar magnetic activity responsible for the corrosive UEX radiation also drives stellar winds and coronal mass ejections (CMEs) that could be even more perilous to atmospheric survival \citep[e.g.][]{Khodachenko.etal:07,Lammer.etal:07}. 

In earlier work, we have applied a state-of-the-art magnetohydrodynamic (MHD) stellar wind and magnetosphere models to begin to investigate the space environment and its atmospheric impact for planets in the habitable zones of active M dwarf stars \citep{Cohen.etal:14,Cohen.etal:15}. These studies introduced generic magnetized and non-magnetized terrestrial planets.  Here, we perform the first step in this process and apply similar stellar wind modeling to estimate the space weather conditions experienced at the orbit of Proxima b. 

\section{Magnetohydrodynamic Modeling}

We simulate the stellar corona and stellar wind using the {\it BATS-R-US} MHD code \citep{Vanderholst.etal:14}. The model is driven at its inner boundary by the stellar surface magnetic field (magnetogram), and it solves the set of non-ideal MHD conservation equations for the mass, density, magnetic induction, and energy.   The model assumes that the coronal heating and wind acceleration are due to Alfv\'en wave turbulence (and thus they depend on the magnetic field strength), and it takes into account radiative cooling and electron heat conduction. The wave dissipation energy is responsible for the plasma heating, while the additional momentum to accelerate the wind is provided by the wave pressure gradient term.  The calculation of the Poynting flux in the model takes into account the observed scaling between the magnetic flux and the activity level \citep{Pevtsov.etal:03}.

\begin{figure}
\center
\includegraphics[trim = 1.1in 0.2in
  0.8in 3.8in,clip, width =  0.45\textwidth]{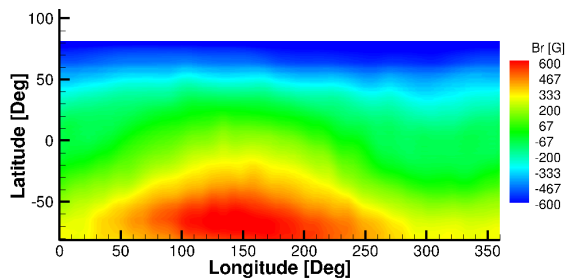} 
\caption{Magnetogram for GJ 51 with $600$~G maximum field strength. }
\label{fig:magnetogram}
\end{figure}

In the case of the Sun, high quality magnetograms used as the boundary conditions to drive the wind model are routinely obtained in exquisite detail. In the stellar case, we have employed surface magnetic field maps derived using the Zeeman-Doppler Imaging (ZDI) technique \citep{Semel:80,Donati.etal:89}.  
The very long rotation period of Proxima Centauri ($\sim 83$ days, \cite{Kiraga.Stepien:07}) renders rotation-based Doppler shifts well below the resolution limit, and no surface magnetic field maps are currently available.  We therefore use the ZDI map of GJ~51 \citep{Morin.etal:10}, an M dwarf of similar spectral type (M5) to Proxima. The difference is that GJ~51 is a much faster rotator (with a rotation period of $\sim 1$~day).  Magnetic field strength scales with rotation period, and to account for this we have scaled the magnetic map to match the field strength measured for Proxima of $\sim$600~G \citep{Reiners.Basri:08}.  Since this is only a single measurement of the average magnetic field strength and magnetic cycles have been detected on Proxima  \citep{Wargelin.etal:16},  we probe two different scalings.  In one, the amplitude (or maximum strength) of the magnetic field is 600~G (see Figure~\ref{fig:magnetogram}) and in the other the mean magnetic field is 600~G and the maximum value is 1200~G (and looks exactly like the magnetogram in Figure~\ref{fig:magnetogram} but with a color bar going up to 1200~G).   

Rotation can also have an impact on the magnetic field geometry, 
with field complexity changing with rotation rate \citep{Vidotto.etal:14b,Reville.etal:15a, Garraffo.etal:15}.  Looking at ZDI maps of other systems with a similar Rossby number to Proxima ($R_o=0.65$; \citealt{Wright.Drake:16}) we find that the magnetic field structure for these systems is relatively simple and similar to a dipole \citep[see, for example, GJ 49 from][]{Morin.etal:08}. Of the stars with available ZDI maps of similar spectral type to Proxima, GJ 51 is has the simplest magnetic field geometry, comparable to the systems with Proxima's Rossby number of $\sim 0.65$.

While the orbital semi-mayor axis of Proxima~b is estimated to be 0.049~AU and its eccentricity is $e <0.3$, its inclination has not been constrained. In order to explore the space of parameters within these limits, we model four orbits: a low ($i = 10$~deg) orbit and a high ($i = 60$~deg) orbit, each with two different eccentricities of $e = 0$ (circular) and $e = 0.2$. Notice that here the inclination refers to that between the orbit of the planet and the rotational axis of the star and not with respect to the observer.

\cite{Zuluaga.Bustamante:16} have estimated that Proxima~Centauri~b is an Earth-like planet with a magnetic moment of about 0.1--1$\mathcal{M}_\earth$ and \cite{Ribas.etal:16} find its obliquity likely to be null. In order to calculate the size of the magnetosphere over the planet's orbit, we explore two cases assuming a magnetic field strength of $0.1~G$, consistent with the range estimated by \citet{Zuluaga.Bustamante:16}, and of $0.3~G$ to match the Earth's present date magnetic moment, both with null obliquity. 

The MHD model of the stellar corona, wind and magnetic field of Proxima, driven using the scaled GJ~51 magnetogram, was computed using {\it BATS-R-US}. We assumed a stellar rotation period of 83~d \citep{Benedict.etal:98}.  Unlike the case of rapidly rotating stars, where rotation results in azimuthal wrapping of the magnetic field even within the Alfv\'en radius \citep{Cohen.etal:10},  for periods longer than twenty days or so the rotation makes very little difference to the magnetospheric structure. We extracted the stellar wind and magnetic field parameters over different plausible orbits based on the constraints of \citet{Anglada-Escude.etal:16}.  In principle, unless the planet is in a retrograde orbit, it takes 12.9 days for the planet to orbit around a given point on the surface of the star. We ignore the effect of stellar rotation here and use orbital phase as the reference frame rather than orbital phase relative to stellar rotational phase. 

We believe the model we have produced is the most realistic simulation one can make at the present time with the available data. As a consistency test we have calculated the mass and angular momentum rates for our wind model and find that the mass loss rate ($\sim 1.5\times 10^{-14}~M_\odot$~yr$^{-1}$) is consistent with the upper limit of  $3\times10^{-13} M_\odot$~yr$^{-1}$ based on the quite direct technique of X-ray charge exchange measurements by \citet{Wargelin.Drake:02}.  However, it is slightly higher than
the upper limit obtained from astrospheric absorption by \citet{Wood.etal:01} of $\sim 0.2 \times 10^{-14}M_\odot$~yr$^{-1}$. \citet{Wood.etal:01} note that their estimate is quite model dependent, and in light of recent research indicating the structure of the heliosphere is very different to the canonical cylindrically symmetric comet-shape \citep[e.g.][]{Opher.etal:15}, we do not consider the discrepancy to be severe. Our model has an angular momentum loss rate of ($\sim 10^{29}$~erg), which for Proxima corresponds to a spin-down timescale of  $\sim 10$~Gyr, consistent with observations and expectations for late M dwarfs \citep{Basri.Marcy:95}. 

\section{Results and Discussion}

\begin{figure*}[][h]
\center
\includegraphics[trim = 1in 2.5in
  0.9in 2.5in,clip, width = 0.48 \textwidth]{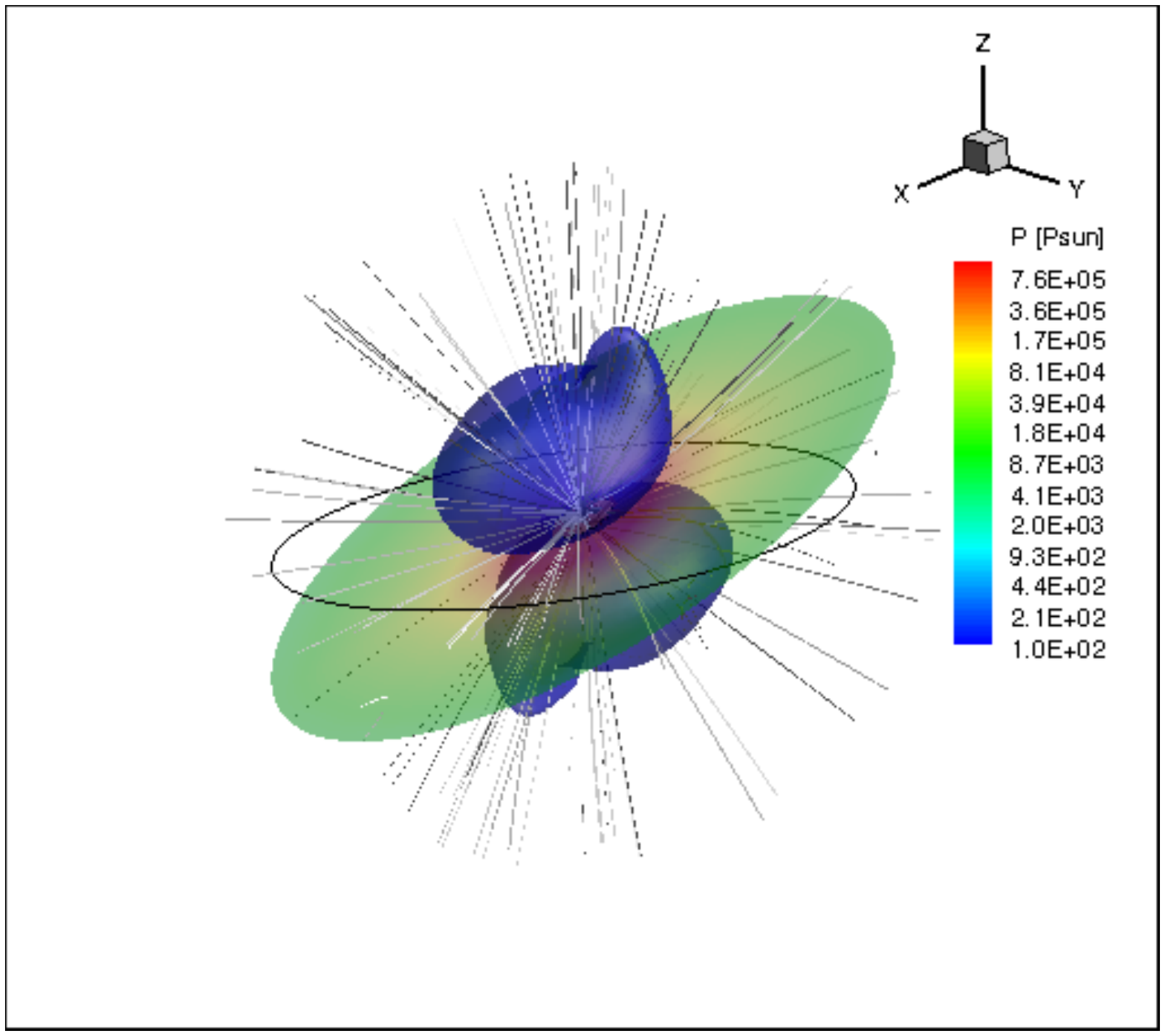} 
\includegraphics[trim = 1in 2.5in
  0.9in 2.5in,clip, width = 0.48 \textwidth]{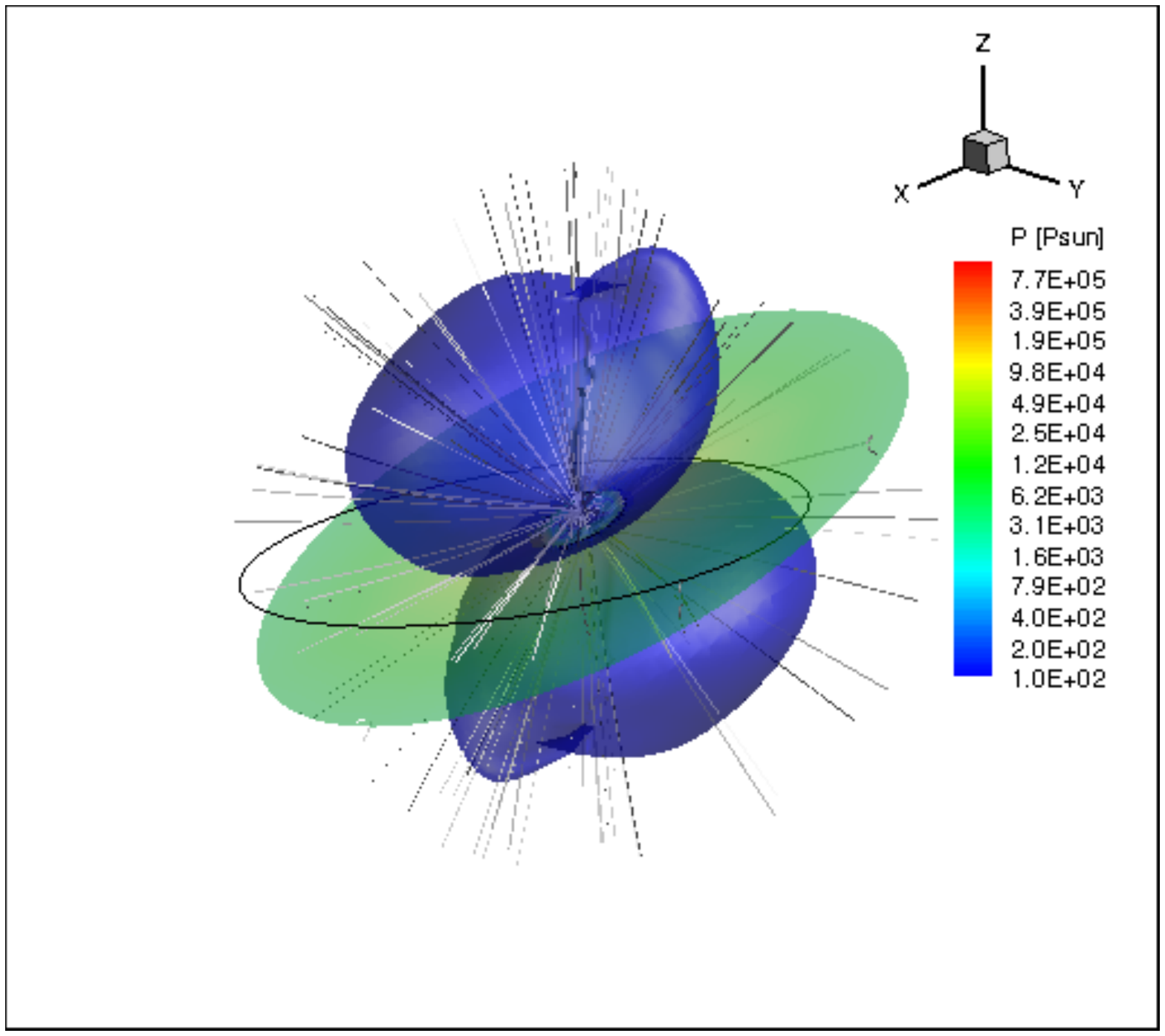} 
\caption{Three-dimensional stellar magnetosphere and wind for Proxima corresponding to a maximum (left) and mean (right) magnetic field of $600$~G (see text).  The Alfv\'en surface is shown in blue and the plane corresponds to the current sheet, which is colored according to the wind dynamic pressure normalized to that of the sun at $1$~AU.  An orbit with Proxima b's semi-mayor axis of $0.049$~AU, $10$~deg inclination, and eccentricity of $0.1$ is shown in black while selected magnetic field lines are gray.}
\label{fig:3d}
\end{figure*}

We illustrate the wind structure resulting from our simulations in Figure~\ref{fig:3d} for each magnetic field scaling: the first one with 600~G field amplitude and the second one with 600~G average field.  The blue surface shows the Alfv\'en surface, while the plane corresponds to the current sheet with a color coding that reflects the dynamic wind pressure $\rho \cdot v^{2}$ normalized to that of the typical solar wind pressure at 1~AU ($\sim 10^{-8} gr/( cm \cdot s^2) $ or 1 nanoPascal). Only one of the modeled orbits is shown illustratively in each plot to avoid confusion, and selected magnetic field lines are plotted in gray. The wind speeds obtained by our model are not dramatically different to the solar wind ones---up to 1300~km~s$^{-1}$ for the lower magnetic field case and up to 1600~km~s$^{-1}$ for the higher magnetic field case compared to 800--900~km~s$^{-1}$ for the fast solar wind \citep{McComas.etal:07}. 
However, the densities at Proxima b's orbital distance are 100 to 1000 times larger than the densities for the solar wind at 1~AU (1--10~cm$^{-3}$). Consequently, the dynamic pressure at the orbital distance of Proxima~b is very high, and three to four orders of magnitude higher than that experienced by the Earth.  The wind dynamic pressure exceeds the magnetic pressure by about an order of magnitude at high latitudes and by up to three orders of magnitude close to the current sheet.

\begin{figure*}[][h]
\center
\includegraphics[trim = 1in 0.2in
  0.9in 0.1in, clip, width = 0.49 \textwidth]{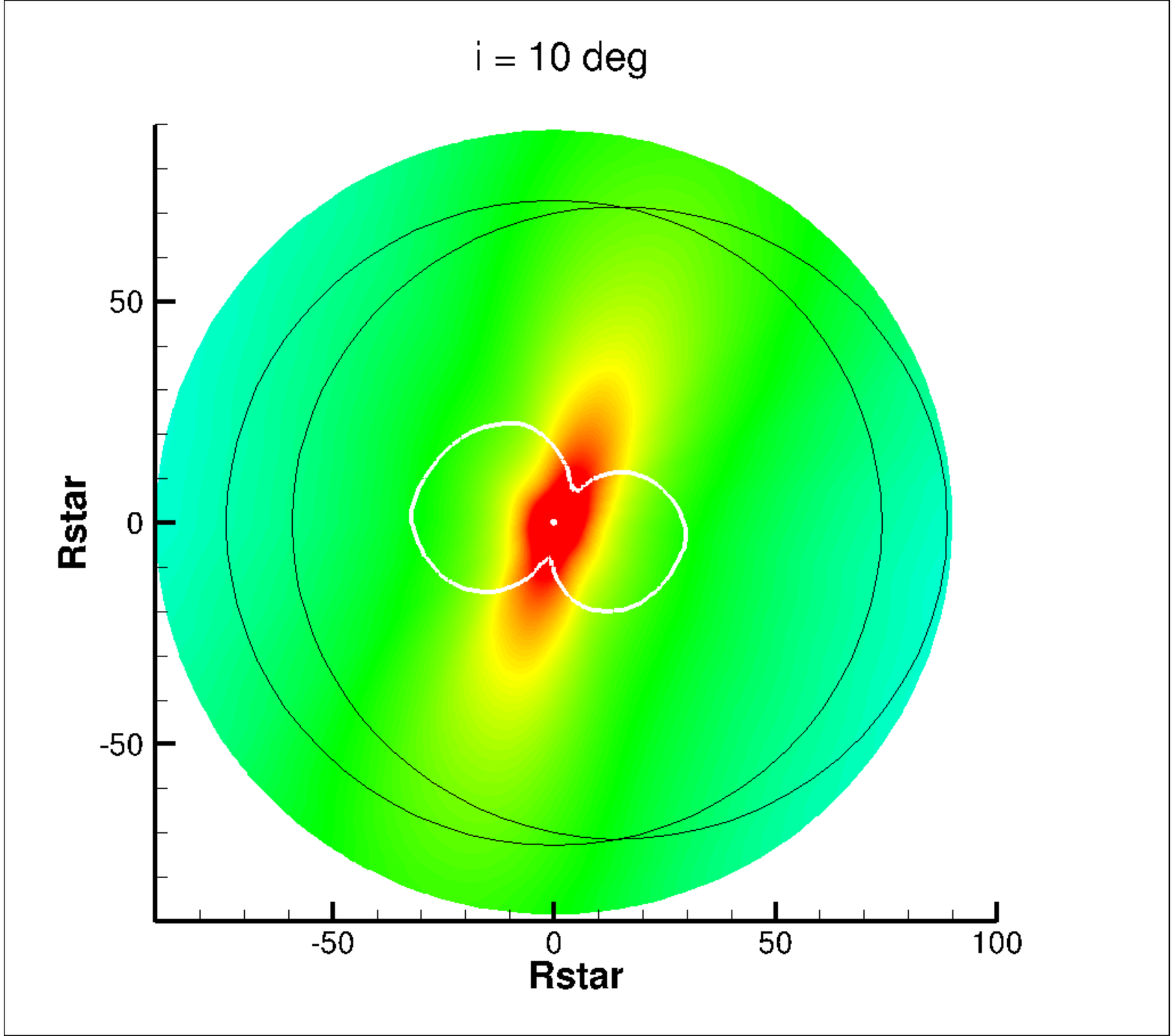}
  \includegraphics[trim = 1in 0.2in
  0.9in 0.1in,clip, width = 0.49 \textwidth]{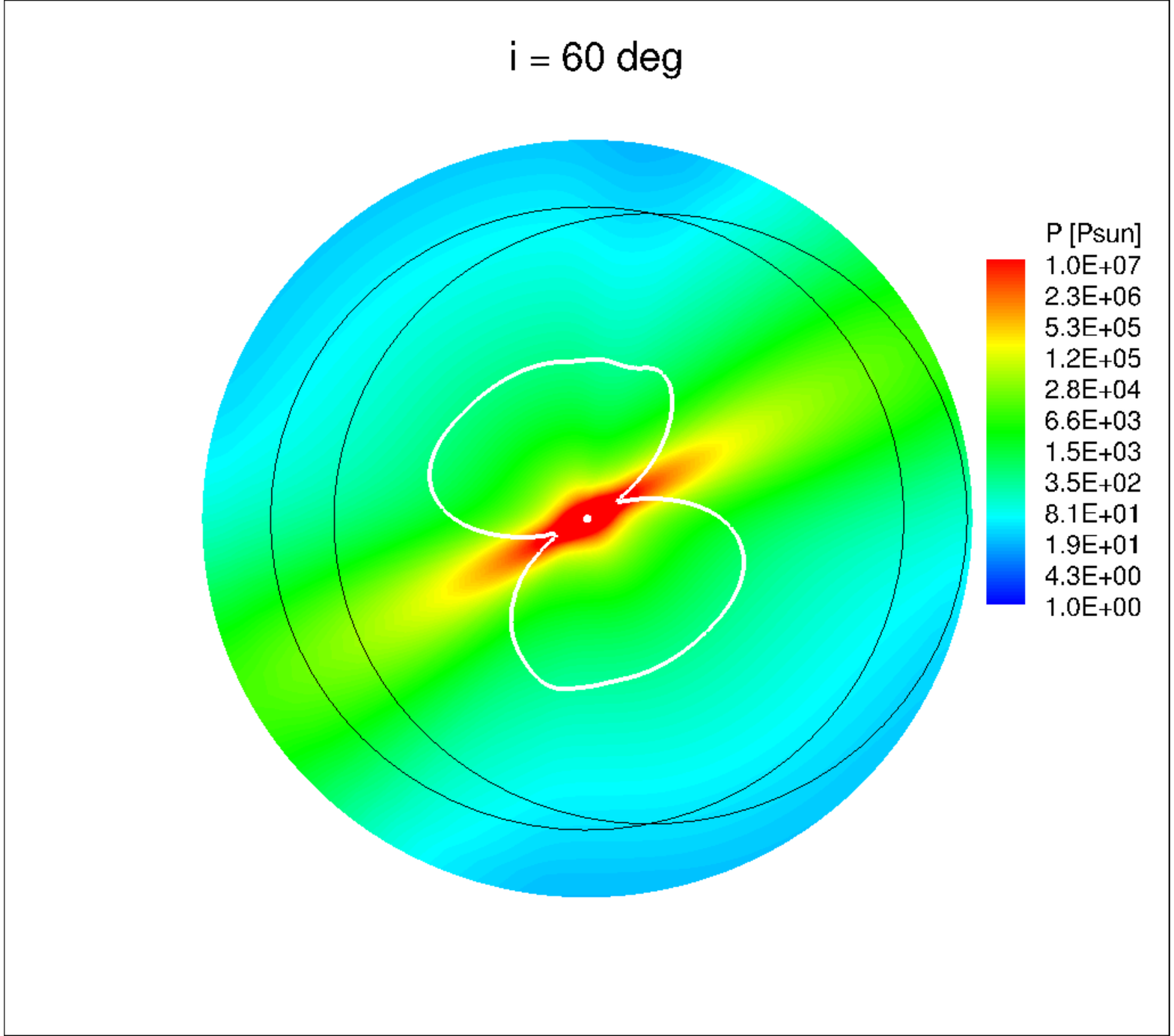}\\
\includegraphics[trim = 1in 0.2in
  0.9in 0.1in,clip, width = 0.49 \textwidth]{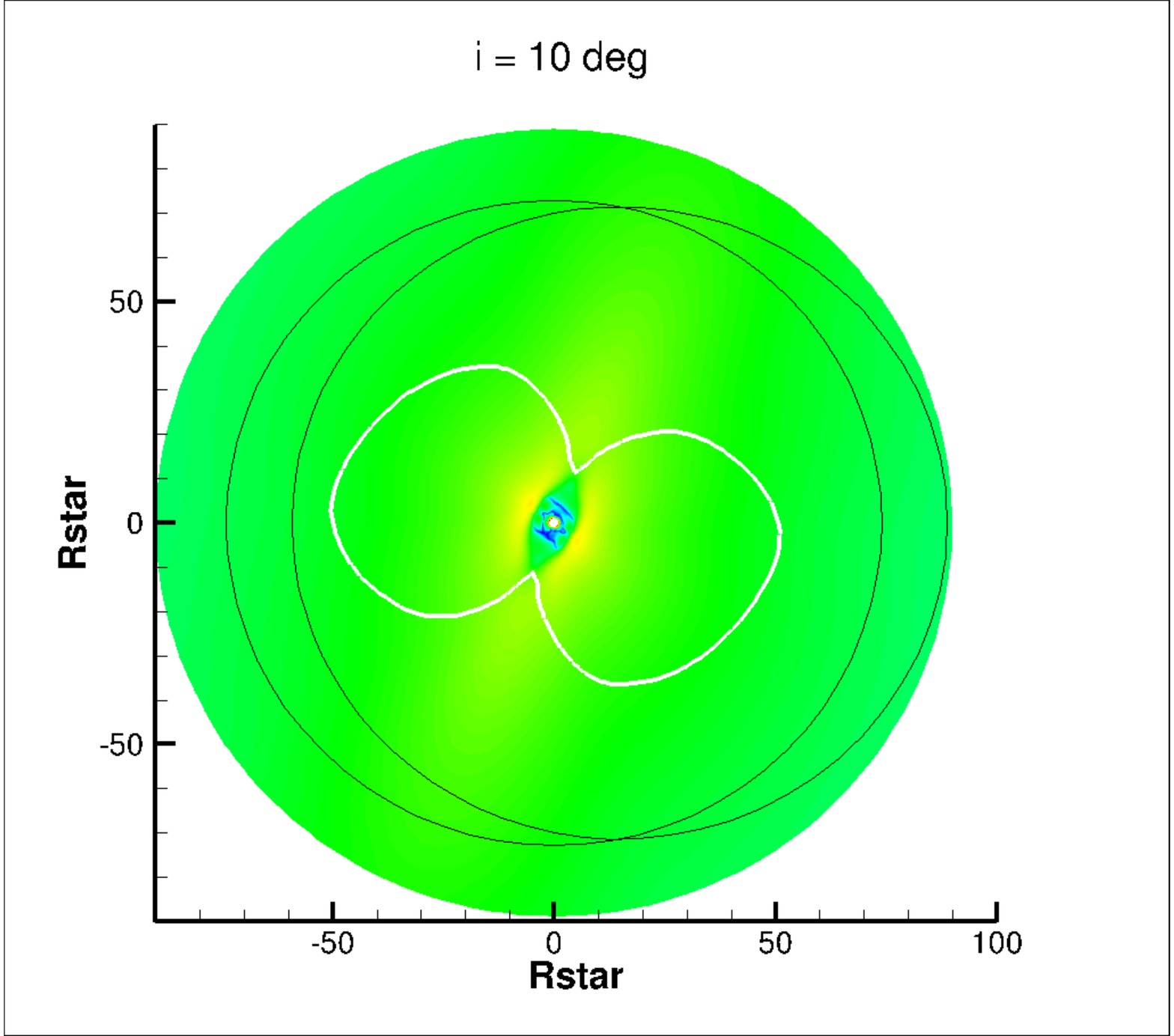}
  \includegraphics[trim = 1in 0.2in
  0.9in 0.1in,clip, width = 0.49 \textwidth]{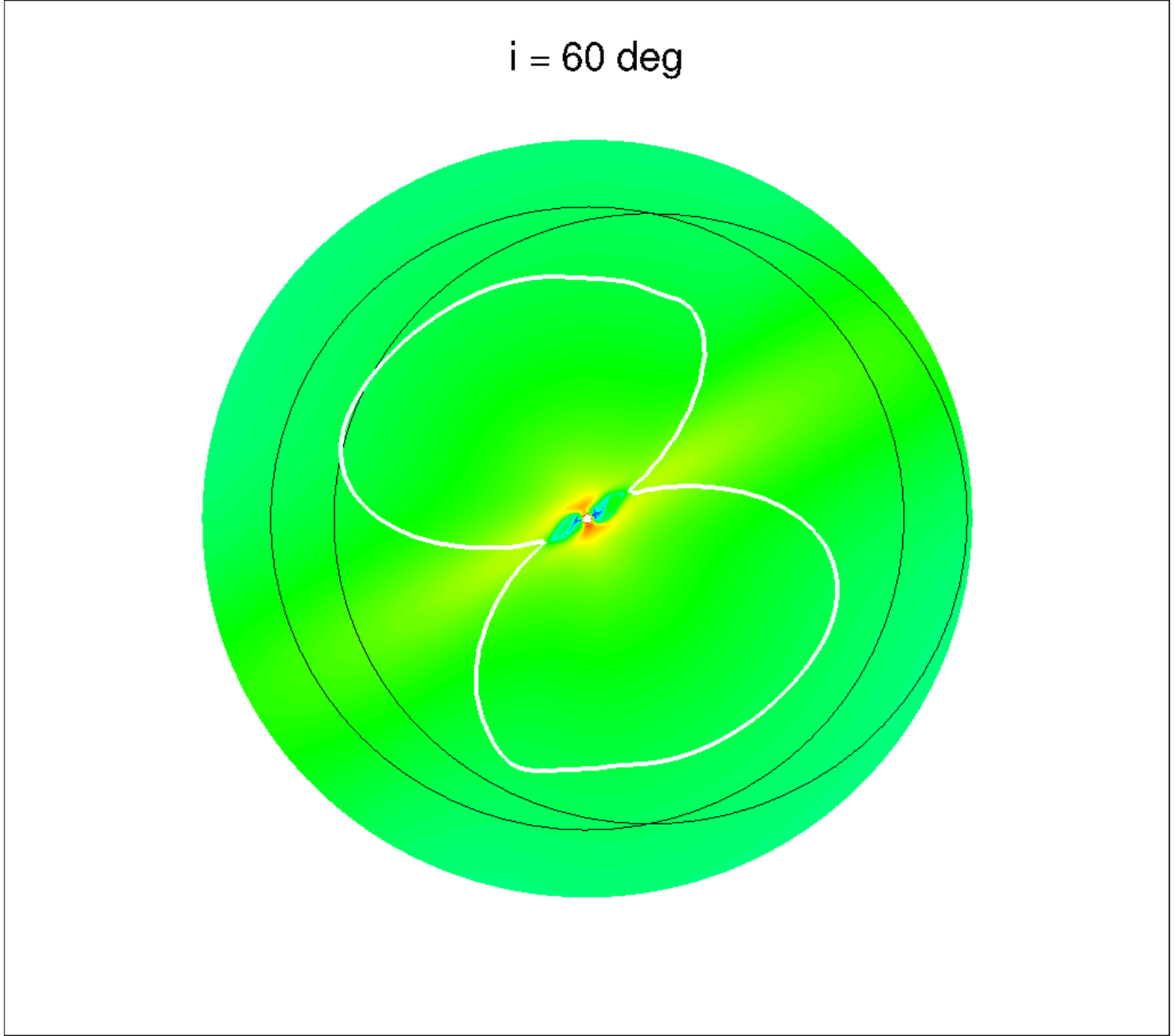}
\caption{Wind pressure normalized to the solar wind at $1$~AU for two different orbits of Proxima~b ($e=0.1, 0.2$) with $i=10$~deg (left panel) and $i=60$~deg (right panel) for a maximum (top panel) and mean (bottom panel) magnetic field of $600$~G. The white line represents the Alfv\'en surface.}
\label{fig:planes}
\end{figure*}

\begin{figure*}
\center
\includegraphics[ width = 0.78\textwidth]{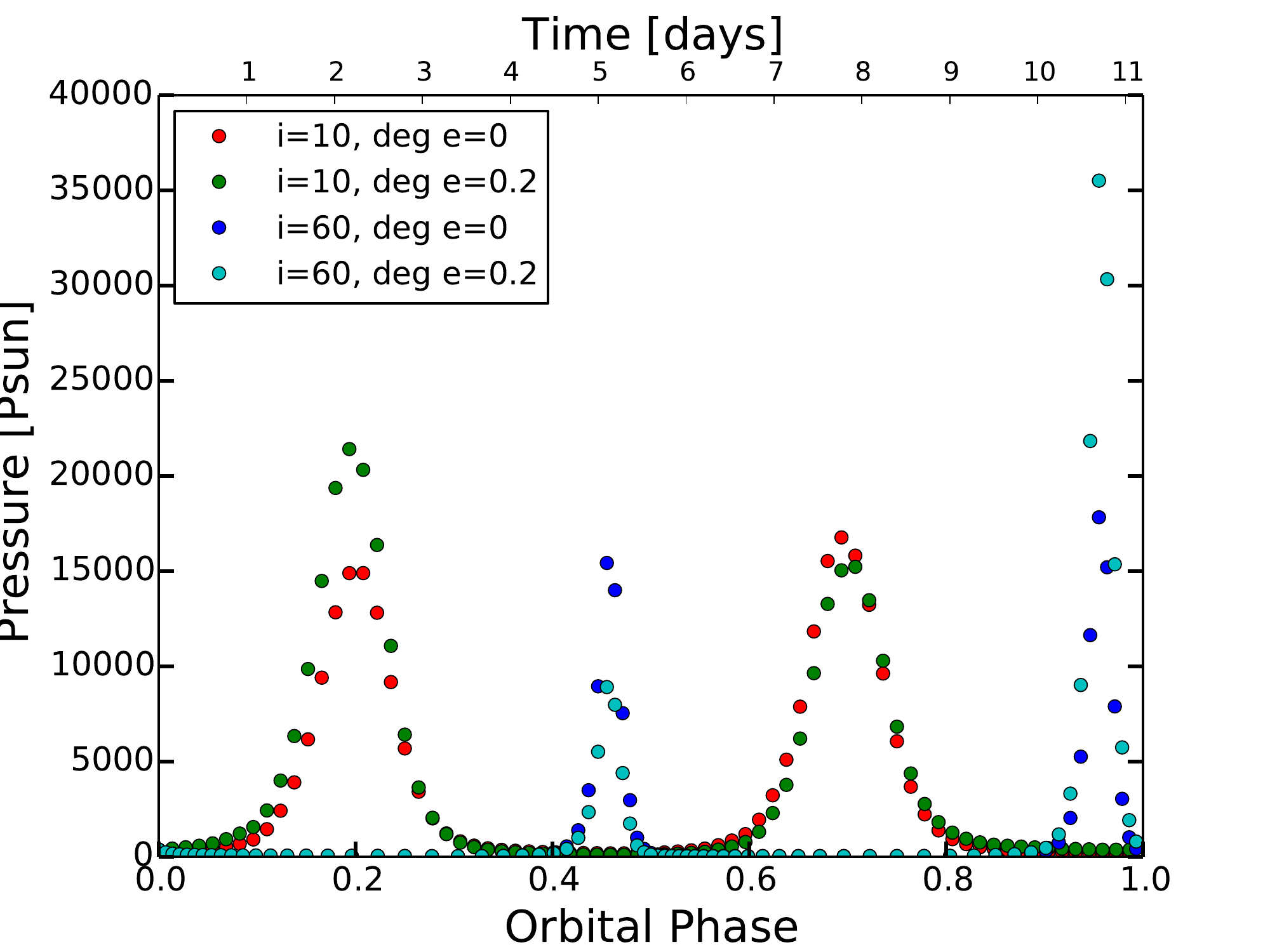}
  \includegraphics[width =  0.78\textwidth]{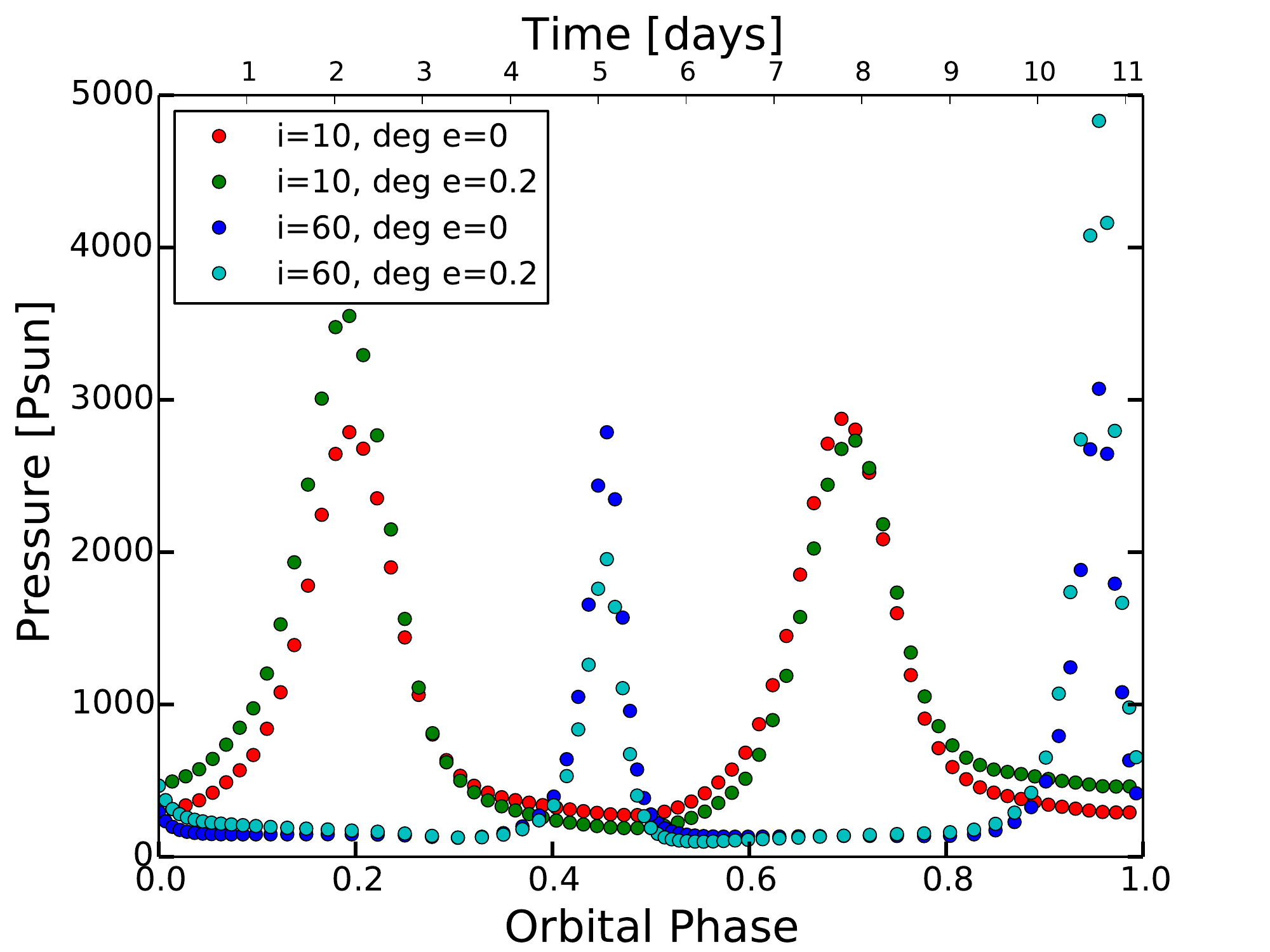}
\caption{Stellar wind pressure for the four modeled orbits ($e=0.1, 0.2$ and $i=10, 60$~deg) for GJ 51 with a maximum (top panel) and mean (bottom panel) magnetic field of $600$~G.}
\label{fig:pressure}
\end{figure*}

Figure~\ref{fig:planes} shows the dynamic wind pressure in the plane of the different orbits considered. In all cases, the planet resides in dynamic wind pressures reaching over a thousand times the one we experience at Earth and passes through large pressure variations over an orbit. In Figure~\ref{fig:pressure} we quantify the pressure as a function of orbital phase for the different orbits examined and for each magnetic field strength.

For the lower magnetic field strength case, all orbits go through a wind pressure change of at least a factor of 1000, while for the stronger magnetic field the variability is smaller but still of at least a factor of 10. In 7 out of 8 cases, the orbits reside close to, but outside of, the Alfv\'en surface.  
For the stronger magnetic field, the orbit of $i = 60$~deg and $e = 0.2$ touches, and maybe crosses, the Alfv\'en surface for a small fraction of its orbital period.  The true Alfv\'en surface will be fairly dynamic and depend on the exact wind conditions at any given time.  The strong dependence of the Alfv\'en surface size on the magnetic field strength and the likely variation of this, together with the recently detected magnetic cycles \citep{Wargelin.etal:16}, suggest Proxima~b is likely to encounter transitions between subsonic and supersonic wind conditions quite frequently.  \citet{Cohen.etal:14} found that the lack of a bow shock in subsonic conditions leads to much deeper penetration of the wind into the magnetosphere than in supersonic conditions. Proxima~b is then likely to experience episodes of deep wind penetration.  While even the Earth can occasionally experience short periods of time under sub-Alfv\'enic wind conditions \citep[e.g.,][]{Chane.etal:15}, the extreme dynamic pressure of the wind conditions for Proxima~b renders the potential effects much more drastic.

\citet{Vidotto.etal:14b} also found changes in the ambient dynamic pressure in their investigation of M~dwarf winds in planetary habitable zones.  However, the dynamic pressure changes we find are orders of magnitude greater than the factor of 3 variations they found. Their wind models were based on an initially spherically-symmetric Parker-type thermally-driven wind \citep{Parker:58} with which the magnetic field subsequently interacts and are quite spatially smooth with densities and pressure variations of factors of only a few.  The models employed here are instead driven by a magnetic field dependent energy deposition that results in a much more spatially variable wind.

The salient effect of the extreme wind pressure at the location of Proxima~b is a compression of its magnetosphere.
As we show below, the magnetopause of Proxima~b lies at a distance between 1 and 5 planetary radii, with average distances likely being 2 to 3 radii.  This is very close to the planet compared with the Earth's magnetopause at about 10 planetary radii, and creates conditions for potentially strong atmospheric stripping by the stellar wind \citep{Cohen:15}. The variations in the magnetopause location due to secular dynamic pressure variations are also expected to drive strong currents in the magnetosphere and ionosphere, and cause heat deposition and particle precipitation down to the upper atmosphere \citep{Kivelson.Russell:95}.

\begin{figure*}
\center
\includegraphics[ width = 0.8\textwidth]{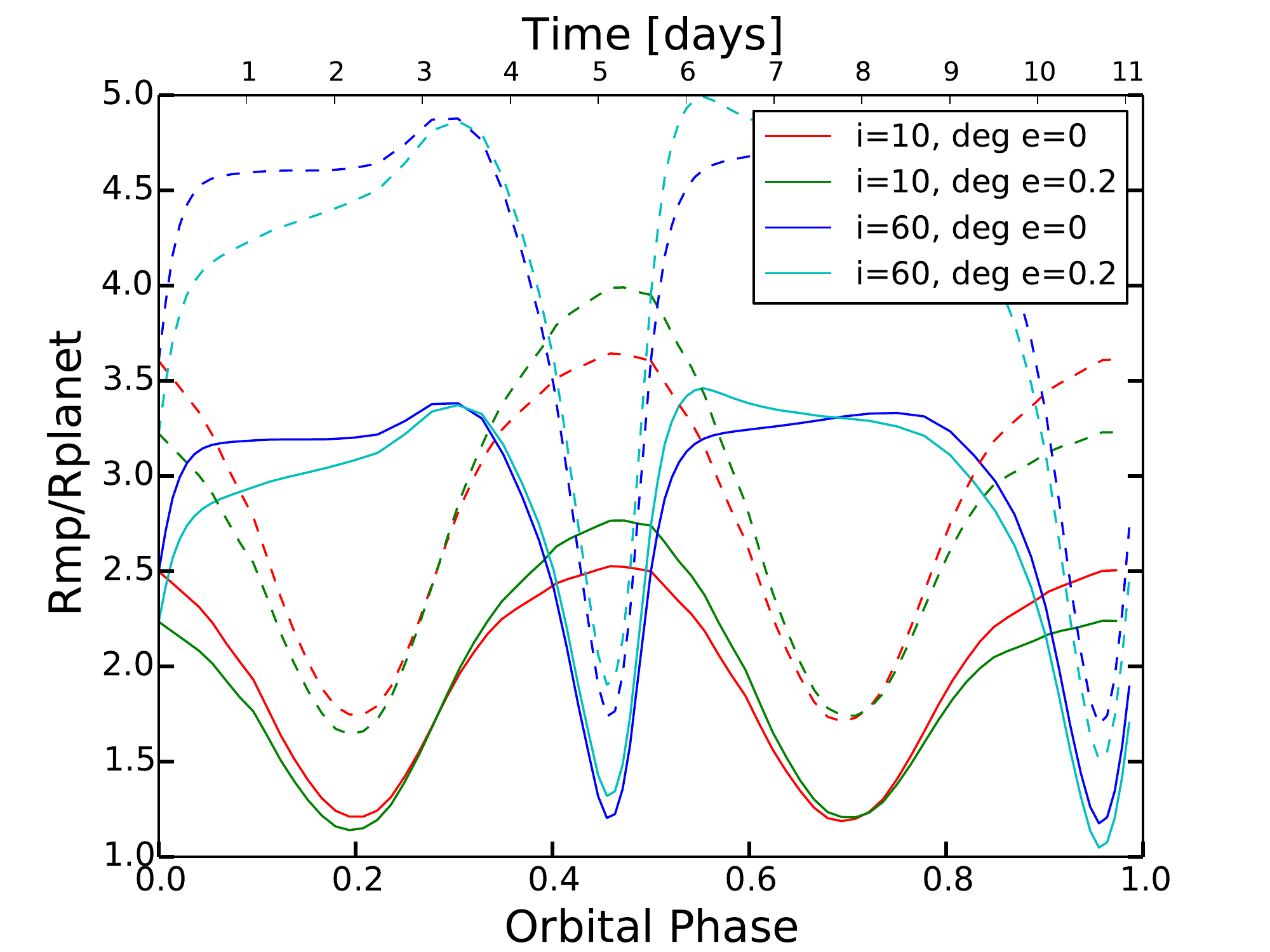}
  \includegraphics[width =  0.8\textwidth]{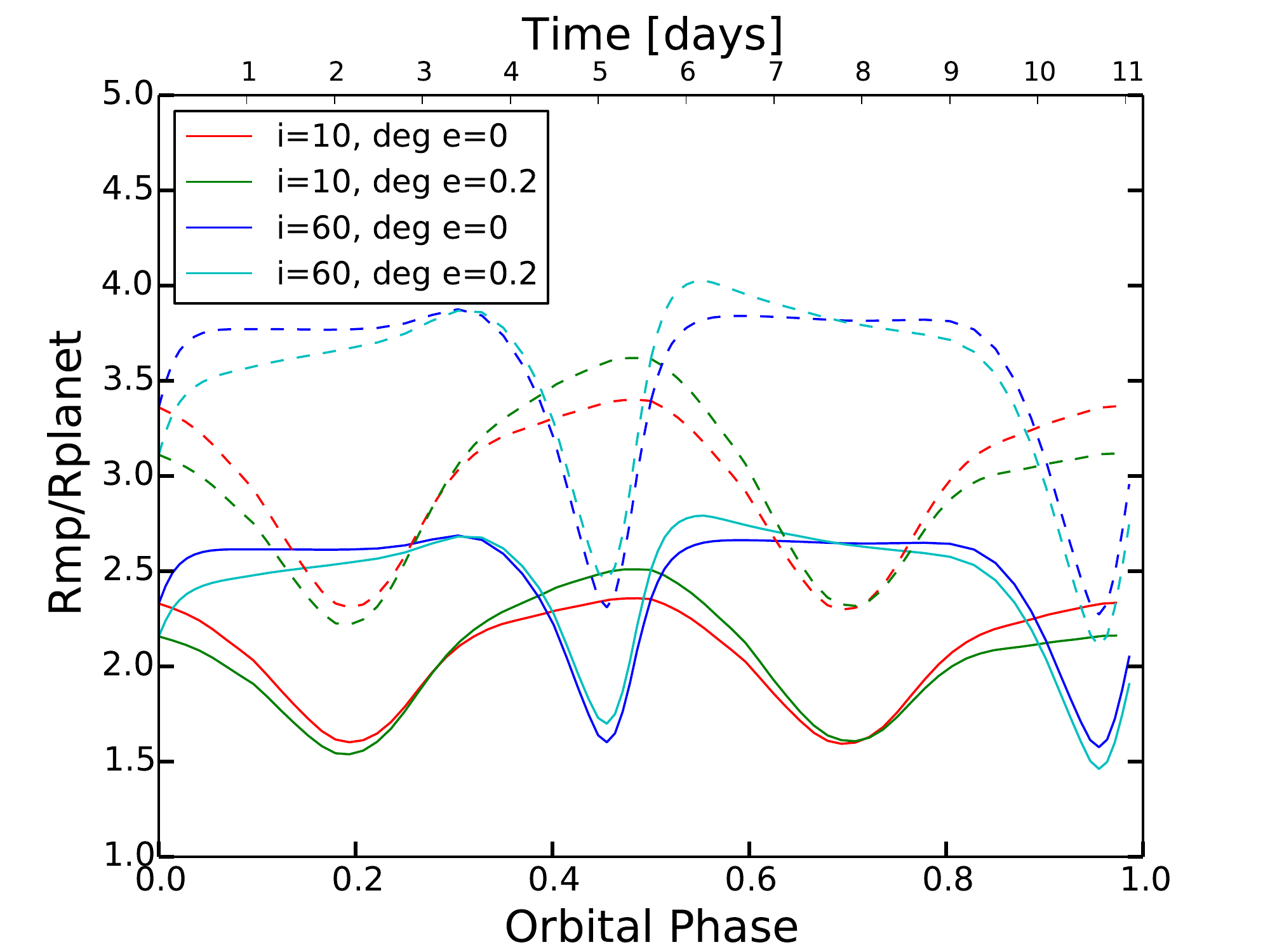}
\caption{Magnetosphere radius normalized to the planet's radius for the four modeled orbits ($e=0.0, 0.2$ and $i=10, 60$~deg for GJ 51 with a maximum (top panel) and mean (bottom panel) magnetic field of $600$~G. The solid line corresponds to a planetary magnetic field like the earth's of $0.3$~G while the dashed line corresponds to a planetary magnetic field of $0.1$~G, expected for Proxima b.}
\label{fig:standoff}
\end{figure*}

In order to better understand how the changes in wind pressure might impact the magnetosphere of Proxima~b, we computed the approximate magnetopause standoff distance as a function of orbital phase for the different orbits considered (see Figure~\ref{fig:standoff}), assuming pressure equilibrium between the stellar wind and the planet's magnetic field \citep[e.g.,][]{Schield:69,Gombosi:04} $$R_{mp}/R_{planet}=[B_p^2/(4\pi P_{SW})]^{1/6},$$ where $R_{mp}$ is the radius of the magnetopause, $R_{planet}$ is the radius of the planet, $B_p$ refers to the planet's equatorial magnetic field strength, and $P_{SW}$ is the ram pressure of the stellar wind. We find that for all the modeled orbits and for both stellar magnetic field scalings, the magnetopause distance changes over the orbit by a factor $>2$ for an Earth-like magnetic field and a factor $> 4$ for a magnetic field of 0.1~G consistent with the \cite{Zuluaga.Bustamante:16} assessment for Proxima~b.  These rapid changes happen twice each orbit as the planet passes through the equatorial streamer regions of dense wind and high dynamic pressure on a timescale as short as a day.  

The magnetopause standoff distance also depends on the strength and orientation of the interplanetary magnetic field relative to that of the planetary field \citep[e.g.][]{Dungey:61,Schield:69,Holzer.Slavin:78}.  The magnetopause is eroded by the transfer of magnetic flux from the day side magnetosphere into the magnetotail through interaction with interplanetary field of opposite polarity.  In the case of the Earth's magnetosphere, \citet{Holzer.Slavin:78} found that flux transfer resulted in magnetopause changes similar to those caused by solar wind pressure variations. Proxima~b is likely to traverse the current sheet twice each orbit (see Figure~\ref{fig:3d}), and the consequent polarity change in the interplanetary magnetic field will add an additional magnetopause change term with a timescale of several days.

\citet{Cohen.etal:14} found that large variations in dynamic pressure like those we find for Proxima~b lead to 
Joule heating at the top of the atmosphere at a level of up to a few percent of the stellar irradiance.
The heating was also enhanced in their time-dependent planetary magnetosphere model because of the additional current generated by the temporal changes in the magnetic field. 
The timescale of the variations for Proxima~b are only of the order of a day. Further detailed and physically realistic modeling is needed in order to asses the full impact of these geomagnetic effects on a planetary atmosphere.

\citet{Wargelin.etal:16} have recently detected a magnetic cycle with a period of 7 years in Proxima.  The nature of the detection in the form of secular changes in $V$ magnitude implies that the cycle modulates the surface starspots.  \citet{Garraffo.etal:15} have shown that spots can close down what would have been open magnetic field, altering the wind mass and angular momentum loss. It is likely that Proxima~b's space environment will then also be subject to associated changes in the magnetic field of its host star.

\cite{Ribas.etal:16} and \citet{Barnes.etal:16} have performed a detailed study based on the orbital evolution and estimated irradiation history pf Proxima~b and concluded that different formation scenarios could lead to a range of situations from dry planets to Earth-like water contents to waterworlds.  \cite{Turbet.etal:16} studied two likely rotation modes (tidally locked and 3:2 resonances) reaching the conclusion that, for both synchronous and non-synchronous rotation, there are scenarios which would allow liquid water to be present. Other studies have aimed at understanding the likely climate, photosynthetic properties and observational discriminants of the atmosphere of Proxima~b \citep{Goldblatt:16,Lopez.etal:14, Meadows.etal:16,Kreidberg.Loeb:16}.  We have shown here that space weather conditions of Proxima~b are extreme and very different to those experienced on Earth. Further work is needed in order to investigate the atmospheric loss processes and rates of Proxima~b and to assess the likelihood that the planet has retained an atmosphere at all.

\section{Summary}

Realistic MHD models of the wind and magnetic field environment of Proxima Centauri show that Proxima~b experiences a stellar wind dynamic pressure three to four orders of magnitude higher than that of the solar wind at Earth.  The pressure is highly nonuniform, being greater in the vicinity of the more dense wind near the current sheet by factors of 10 to 1000, depending on the stellar magnetic field.  Proxima~b will pass through these extreme pressure variations twice each orbit, leading to magnetospheric compression and expansion by a factor up to 3 on a timescale of about a day. Its orbit is also likely to pass in and out of the Alfv\'en surface, thus exposing it to both subsonic and supersonic wind conditions.  These phenomena are expected to have a significant effect on any atmosphere of Proxima~b.

\acknowledgments

We thank anonymous referee for very helpful comments.  CG was supported by SI Grand Challenges grant ``Lessons from Mars: Are Habitable Atmospheres on Planets around M Dwarfs Viable?''.  JJD was supported by NASA contract NAS8-03060 to the {\it Chandra X-ray Center}.  OC was supported by NASA Astrobiology Institute grant NNX15AE05G. Simulation results were obtained using the Space Weather Modeling Framework, developed by the Center for Space Environment Modeling, at the University of Michigan with funding support from NASA ESS, NASA ESTO-CT, NSF KDI, and DoD MURI. Simulations were performed on NASA's PLEIADES cluster under award SMD-16-6857.  CG thanks Rakesh K. Yadav for useful comments and discussions.


\end{document}